# Electron Beam Restructuring of Quantum Emitters in Hexagonal Boron Nitride

**Sergei Nedić[1,2†], Karin Yamamura[1,2†], Angus Gale[1,2], Igor Aharonovich[1,2] and Milos Toth[1,2]**

[1] *School of Mathematical and Physical Sciences, Faculty of Science, University of Technology Sydney, Ultimo, New South Wales 2007, Australia*

[2] *ARC Centre of Excellence for Transformative Meta-Optical Systems (TMOS), University of Technology Sydney, Ultimo, New South Wales 2007, Australia*

*† These authors contributed equally to this work.*

angus.gale@uts.edu.au; milos.toth@uts.edu.au

## Abstract

Hexagonal boron nitride (hBN) holds promise as a solid state, van der Waals host of single photon emitters for on-chip quantum photonics. The B-centre defect emitting at 436 nm is particularly compelling as it can be generated by electron beam irradiation. However, the emitter generation mechanism is unknown, the robustness of the method is variable, and it has only been applied successfully to thick flakes of hBN (>> 10 nm). Here, we use in-situ time-resolved cathodoluminescence (CL) spectroscopy to investigate the kinetics of B-centre generation. We show that the generation of B-centres is accompanied by quenching of a carbon-related emission at ~305 nm and that both processes are rate-limited by electromigration of defects in the hBN lattice. We identify problems that limit the efficacy and reproducibility of the emitter generation method, and solve them using a combination of optimized electron beam parameters and hBN pre- and post-processing treatments. We achieve B-centre quantum emitters in hBN flakes as thin as 8 nm, elucidate the mechanisms responsible for electron beam restructuring of quantum emitters in hBN, and gain insights towards identification of the atomic structure of the B-centre quantum emitter.

## Introduction

Solid state single photon emitters (a.k.a. "quantum emitters") are a key component of emerging quantum technologies.[1,2] Hexagonal boron nitride (hBN), a van der Waals dielectric with a wide bandgap of ~6 eV,[3] has gained interest as a material host of quantum emitters because it features a wide range of stable fluorescent defects.[4–7] In particular, the B-centre defect has recently gained attention because it has a zero-phonon line (ZPL) that is positioned consistently at 436 nm[5,8,9], and it can be generated on demand using site-specific electron beam irradiation.[10] These properties make it appealing for integrated on-chip quantum photonics due in part to ease of incorporation in devices such as photonic crystal cavities.[11]

However, robustness of the electron beam B-centre generation technique is poor, and it has to date been applied successfully only to thick flakes of hBN (>> 10 nm). Interestingly, it has been noted[10] that the technique works only with hBN that contains a fluorescent defect with a ZPL at 305 nm,[4] which has been correlated with the presence of carbon in hBN.[12–16] Carbon has therefore been postulated to play a role in the atomic structure of the B-centre defect.[17] However, this is a matter of debate. Alternatives have been proposed such as the split nitrogen interstitial defect,[17,18] but the atomic structure of the B-centre remains unknown. Moreover, the mechanism by which B-centres are generated by electron beam irradiation is also unknown, and the reproducibility and efficiency of the method are variable for reasons that are not well understood.

Here we study the B-centre defect using in-situ cathodoluminescence (CL) spectroscopy. The focused electron beam of a scanning electron microscope (SEM) is used to simultaneously modify hBN and excite CL emissions. Time-resolved CL spectroscopy reveals that the B-centre and UV defects are both restructured by the beam. Specifically, the B-centre and UV emission intensities grow and decay during electron irradiation, respectively, at rates that are different for each emission. The measured CL kinetics reveal that the 305 nm UV defects are not converted to B-centres through a simple one-step process such as impact ionization or cluster decomposition. Instead, the B-centre generation and UV quenching are both rate-limited by mass transport (i.e., electromigration) of charged defects within the hBN lattice. The rate of electromigration is influenced by two key factors – the rate at which electrons scatter at defects in hBN, and the intensity of an electric field generated in hBN due to charging. Understanding these mechanisms allows us to optimize the electron beam irradiation parameters, and the measured CL kinetics provide insights towards identification of the atomic structure of B-centre quantum emitters.

In addition, we also show that electron beam irradiation causes deposition of thin films on the surface of hBN due to immobilization of hydrocarbon contaminants. Whilst this phenomenon is well known in SEM,[19] we show here that the films interfere with B-centre generation by reducing the amount of energy deposited into hBN by the electron beam. The films also produce a broadband photoluminescence (PL) emission which reduces the signal-to-background ratio of B-centre quantum emitters. These problems can be alleviated using in-situ and ex-situ hBN processing treatments which greatly improve the robustness of the electron beam method, and enable the generation of B-centre quantum emitters in thin flakes of hBN (< 10 nm).

## Results and Discussion

### Restructuring of defects by electron irradiation

A schematic of the electron beam irradiation setup is shown in Figure 1(a). A standard SEM was used to irradiate carbon-doped hBN[20] with an electron beam, which simultaneously modifies defects in hBN and excites CL, as is depicted in Figure 1(b). The emitted light was collected using a ball lens coupled to an optical fibre connected to a spectrometer located outside the SEM vacuum chamber. To improve the CL collection efficiency, the sample was tilted to 45°, as is shown in Figure 1(a,b). All CL measurements are from ensembles of quantum emitters. In some experiments, a capillary-style gas injection system (GIS) was used to inject $H_2O$ vapor into the SEM chamber during electron irradiation.[21] A number of hBN flakes were employed for CL measurements, with thicknesses in the range of ~75 nm to ~200 nm. Unless stated otherwise, the electron range was greater than the hBN thickness, as is illustrated in Figure 1(b), where the black arrow represents the electron beam and the dashed red teardrop represents the electron-solid interaction volume. PL measurements were performed *ex-situ* using a 405 nm continuous wave laser, as is detailed in the Methods section. In all samples, the B-centre emission was absent in PL spectra acquired prior to electron irradiation.

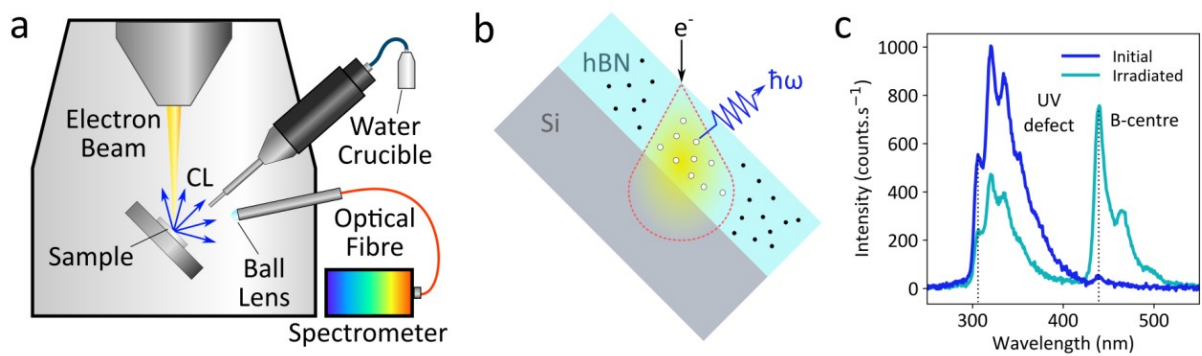

*Figure 1.* Experimental setup used for electron irradiation of hBN and for CL spectroscopy. (a) Schematic of the SEM used in this work. (b) Close-up showing the electron interaction volume (dashed red line) in a flake of hBN on a Si substrate. Defects in hBN (black and white circles) within the interaction volume (white circles) are modified by the beam, and CL is emitted from defects excited by the electrons. (c) CL spectra showing the UV and B-centre emissions at the start and end of a 300 s electron beam irradiation treatment. The UV and B-centre ZPLs are indicated by vertical dashed lines. The UV and B-centre emission intensities decrease and increase, respectively, as a result of the electron irradiation. [Spectrum integration time = 2 s, electron beam energy = 5 keV, beam current = 1.6 nA.]

Typical CL spectra acquired from hBN at the start and end of a 300 s electron beam (5 keV, 1.6 nA) irradiation treatment are shown in Figure 1(c). The UV emission consists of a ZPL at ~305 nm, and phonon replicas at ~320, 334 and 351 nm.[22] The B-center emission consists of a ZPL at 436 nm and two phonon replicas at ~462 and 491 nm.[23] The CL spectra, labeled "initial" and "irradiated", illustrate the effect of the 300 s electron beam irradiation on the emissions. The B-centre emission is negligible in CL spectra acquired at the start of the electron beam irradiation treatment (a very weak emission is seen in the figure due to the 2 s irradiation needed to acquire the initial spectrum). During irradiation, the intensity of the UV emission decreases and that of the B-centre increases with time. The changes in spectra are irreversible.

The spectra in Figure 1(c) show that electron irradiation can cause both the quenching and generation of CL emissions in hBN, as has been demonstrated previously in other materials that include diamond, $SiO_2$ and GaN.[24–29] To gain insights into the underlying mechanisms, we used the following procedure. CL spectra were collected continuously at 2 s intervals for a total electron beam irradiation time of 300 s. The UV and B-centre emission intensities in each spectrum were integrated over the spectral ranges shown in Figure 2(a), and the intensities were plotted versus time, as is done in Figure 2(b) and (c). The experiment was repeated on pristine regions of hBN versus electron beam parameters such as the beam current (Figure 2(a-c)) and energy (Figure 2(d-f)). To quantify and compare the CL kinetics, the time-resolved B-center and UV data were fit using Equations 1 and 2, respectively:

$$I = I_{inf}\left(1 - e^{-t/t_{sat}}\right) \qquad (1)$$

$$I = I_{init} - I_{inf}\left(1 - e^{-t/t_{sat}}\right) \qquad (2)$$

where $I$ is CL emission intensity, $t$ is electron beam irradiation time, $I_{init}$ is initial CL intensity at $t = 0$, $I_{inf}$ is the CL saturation intensity in the limit $t \to \infty$, and $t_{sat}$ is CL saturation time. The rate of change of CL intensity is therefore quantified by $1/t_{sat}$. Fits obtained using Equations 1 and 2 are shown as solid black lines and $t_{sat}$ is specified in the legend of each CL kinetics plot.

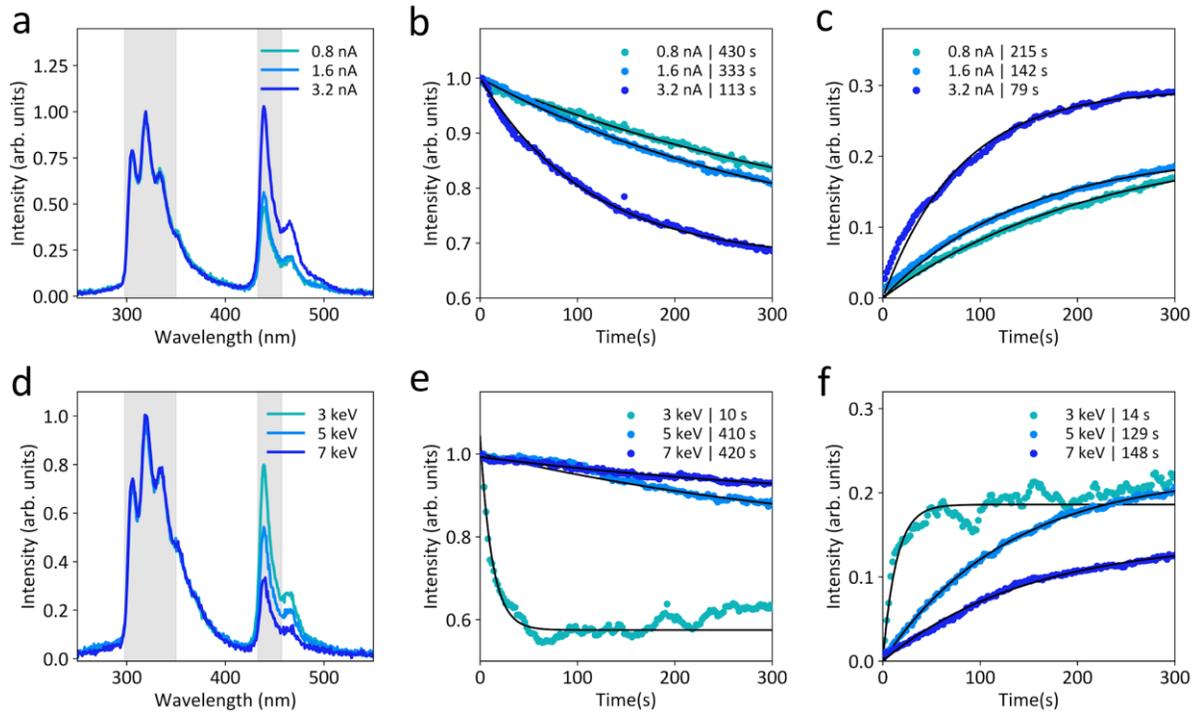

**Figure 2.** CL kinetics versus electron beam current and energy. (a) CL spectra from three regions of hBN collected after each was irradiated for 300 s using a 5 keV electron beam, and beam currents of 0.8, 1.6

*and 3.2 nA. (b) UV emission intensity versus electron beam irradiation time, at each beam current. (c) Corresponding B-centre emission intensity versus time, at each beam current. (d) CL spectra from three regions of hBN collected after each was irradiated for 300 s using electron beam energies of 3, 5 and 7 keV (beam current = 1.6 nA). (e) UV emission intensity versus time, measured at each beam energy. (f) Corresponding B-centre CL intensity versus time, at each beam energy. In all cases, the rate of change of CL intensity increases with current and decreases with electron beam energy. [The spectrum integration time was 2 s. Each spectrum in a and d is normalised to the intensity of the first PSB of the UV emission. Each curve in b, c, e and f is normalised to the CL intensity at zero time. Circles represent experimental data and black lines are fits obtained using Equations 1 and 2. Saturation times, $t_{sat}$, extracted from the fits are shown in the legends.]*

Figures 2(a-c) show the UV and B-centre CL kinetics measured using electron beam currents of 0.8, 1.6 and 3.2 nA, and a fixed electron beam energy of 5 keV. The UV emission CL quenching rate (i.e., $1/t_{sat}$) increases with current, as is indicated by $t_{sat}$ in the legend of Figure 2(b). That is, $t_{sat}$ decreases from 430 s to 113 s, as the beam current is increased from 0.8 to 3.2 nA. Similarly, the B-centre CL generation rate increases with current (Figure 2(c)). However, at each beam current, $t_{sat}$ is consistently smaller for the B-centre emission than for the UV emission. For example, at 0.8 nA, $t_{sat}$ is 430 s and 215 s for the UV and B-centre emissions, respectively. This indicates that the relationship between the UV centre quenching and B-centre generation is not a consequence of competitive recombination (a well-known artifact of CL spectroscopy[30,31]). It also indicates that the underlying physical mechanisms responsible for the CL kinetics are rate-limited by two distinct physical processes. That is, the UV defects are likely not being converted to B-centres in a simple one-to-one process such as impact ionization of the UV defect or decomposition of a defect cluster responsible for the UV emission into B-centre defects. In such processes, $t_{sat}$ is expected to be the same for the UV and B-centre emissions, which is not observed experimentally.

We also note that the $t_{sat}$ values in Figure 2(b,c) do not scale linearly with electron beam current, and that $t_{sat}$ is, in all cases, on the order of minutes even though the electron beam current is on the order of 1 nA (i.e., ~ 6 x 10$^9$ electrons/s). We attribute the nonlinear scaling with current to charging,[32–35] which is discussed in detail below. The long time constants, $t_{sat}$, suggest that the kinetics are rate-limited by a slow process such as electromigration (i.e., mass transport) of defects in the electron interaction volume.[32,34–36] To explore this hypothesis further, we measured CL kinetics as a function of electron beam energy, which is known to influence charging and hence the rate of defect electromigration in wide bandgap dielectrics such as hBN.

Figures 2(d-f) show the UV and B-centre CL kinetics measured at electron beam energies of 3, 5 and 7 keV, using a fixed current of 1.6 nA. The saturation times increase with energy, and the scaling is superlinear. That is, the rate of change of the CL intensity decreases with increasing electron beam energy for both emissions. For the UV emission, $t_{sat}$ ~ 10, 410 and 421 s, whilst for the B-centre emission $t_{sat}$ ~ 14, 129 and 148 s at E = 3, 5 and 7 keV, respectively. This suggests that the CL kinetics are influenced by charging of hBN. To elucidate this point consider the simulated electron energy loss curves shown in Figure 3(a). The plot shows the energy loss rate ($\partial E/\partial z$) of electrons incident on a 125 nm flake of hBN on bulk Si (i.e., the sample used in Figure 2(d-f)), simulated[37] as a function of depth ($z$) below the hBN surface, for electron beam energies of 3, 5 and 7 keV. Integrating each curve between 0 and 125

nm gives the mean energy deposited into hBN per electron, yielding 2.7, 2.6 and 2.1 keV at beam energies of 3, 5 and 7 keV, respectively. That is, the total energy dissipated by the electrons in hBN decreases with increasing beam energy. The reduction is small, but the electron energy density ($\varepsilon$) – i.e., the energy deposited into hBN divided by the electron interaction volume – is large because the interaction volume shrinks rapidly with decreasing beam energy. This is illustrated by the maximum electron range $Z$ (indicated by arrows in Figure 3(a)): $Z \sim$ 111, 247 and 440 nm at electron beam energies of 3, 5 and 7 keV, respectively. Hence, approximating the interaction volume by a hemisphere,[30,31] $\varepsilon \sim$ 9 x $10^{-4}$, 1 x $10^{-4}$ and 3 x $10^{-5}$ eV/nm$^3$, at beam energies of 3, 5 and 7 keV. That is, $\varepsilon$ increases with decreasing electron beam energy in a manner that is highly non-linear, as do the rates of change ($1/t_{sat}$) of the UV and B-centre CL emissions. This is consistent with electromigration, as we explain below.

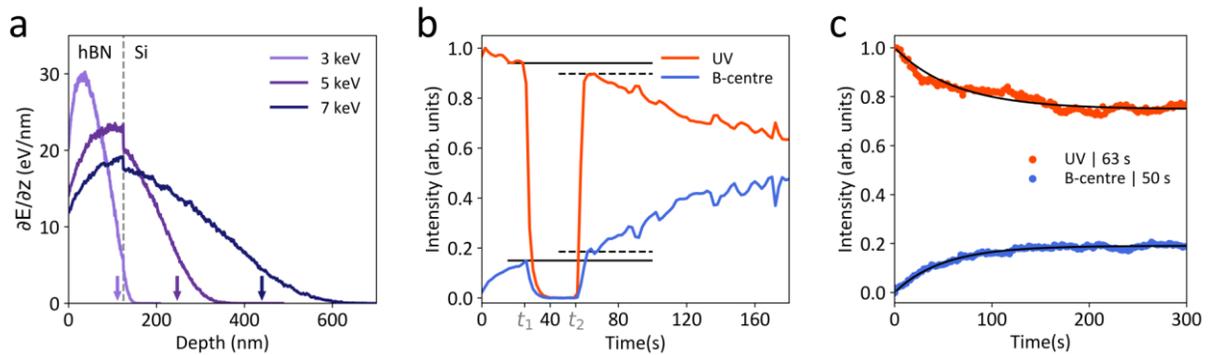

*Figure 3.* Role of energy density $\varepsilon$ and the electric field $\xi$ in CL kinetics. (a) Simulated depth distributions of the energy loss rates ($\partial E/\partial z$) of electrons incident on a 125 nm flake of hBN on bulk Si, calculated for electron energies of 3, 5 and 7 keV. The range (Z), indicated by arrows, is defined here as the mean depth (z) at which 95% of the energy has been dissipated. (b) UV and B-centre CL intensity measured as a function of electron beam irradiation time. The irradiation was interrupted at time $t_1$ for 30 s and resumed at $t_2$. The solid and dashed black lines show the CL intensities at $t_1$ and $t_2$, respectively. The intensities continue to change while the irradiation is interrupted, because $\xi$ does not dissipate instantaneously when the irradiation is interrupted. (c) UV and B-centre CL intensity versus electron beam irradiation time, measured from a hBN flake coated with a 7 nm grounded gold film. The coating causes an increase in the CL kinetics rates, indicated by the short values of $t_{sat}$ specified in the legend. [The electron beam energy and current were 5 keV and 1.6 nA. In (b) the beam was defocused to a diameter of ~40 nm to minimize the effect of errors in the beam placement accuracy when the irradiation was resumed at $t_2$.]

Electromigration is the drift of charged defects under the influence of an electric field ($\xi$).[32,34,35] The electron beam plays two primary roles in the process. First, it generates $\xi$ due to charging of a dielectric, which we discuss below. Second, the product of $\varepsilon$ and the electron beam current, $R = \varepsilon I$, makes defects susceptible to migration because $R$ is proportional to the rate at which electrons scatter with defects (i.e., white circles in Figure 1(b)) in the hBN lattice. The inelastic scattering rate is proportional to the rate at which chemical bonds that bind defects to the lattice are broken, and also to the rate at which the defects are ionized by the beam. The rate of electromigration (quantified by $1/t_{sat}$ in CL data) is therefore expected to scale with $R$, consistent with the results in Figure 2 – i.e., $1/t_{sat}$ is expected to increase with $I$ (Figure 2(b,c)) and $\varepsilon$ (i.e., reciprocal electron beam energy, Figure 2(e,f)). However, whilst consistent, this alone is not proof of electromigration. We therefore now turn to more nuanced experiments focused on the role of $\xi$ in the observed CL kinetics.

First, we note that the electron beam energy affects $Z$ (indicated by arrows in Figure 3(a)), and hence the fraction of charges that can leak to ground efficiently through the grounded Si substrate. The magnitude of $\xi$ and the rate of electromigration are therefore expected to decrease with electron beam energy, consistent with the data in Figure 2(e,f) – i.e., the measured values of $1/t_{sat}$ decrease with increasing beam energy. However, more direct evidence for the role of $\xi$ in the CL kinetics can be obtained by blanking the electron beam during the irradiation process. When the beam is blanked, charged defects can, in principle, continue to migrate for some time determined by the rate at which $\xi$ decays (i.e., the rate at which excess charge dissipates from hBN via leakage currents), the rate at which bonds reconstruct and the rate at which defect charge states recover to equilibrium values determined by the lattice chemical potential. We therefore performed beam blanking experiments exemplified by Figure 3(b), where electron beam irradiation was initiated at 0 s, the beam was blanked at $t_1$ for 30 s, and the irradiation was resumed at $t_2$. Both the UV and B-centre CL profiles show that pausing the irradiation does not interrupt fully the processes responsible for the observed CL kinetics. Instead, at time $t_2$, the UV emission intensity is lower and the B-centre emission intensity is greater than their respective values at $t_1$, as is indicated by solid and dashed black lines in Figure 3(b). This is consistent with a process driven by the electric field $\xi$, and serves as evidence for the role of electromigration in the observed CL kinetics. Moreover, the result in Figure 3(b) eliminates conclusively the possibility of mechanisms that are purely collision-based (i.e., due to electron scattering), since these are interrupted fully while the irradiation is interrupted.

To further substantiate the proposed role of $\xi$ in the observed CL kinetics, we performed one more experiment designed to increase the electric field intensity $|\xi|$. This was achieved by coating a flake of hBN using an electrically grounded thin, metallic film.[32,34,35] $|\xi|$ is determined by the amount of excess trapped charge in hBN generated by the electron beam, and the distance ($s$) over which the resulting potential field ($V$) is terminated (i.e., $|\xi|$ = -d$V$/d$s$). Below the hBN flake, the field terminates at the grounded Si substrate (see Figure 1(b)), whilst above the hBN, it terminates at the closest ground plane in the SEM vacuum chamber (e.g., one of the grounded metal objects shown in Figure 1(a)). The grounded coating terminates $V$ at the hBN surface and therefore reduces $s$ and increases $|\xi|$ within the hBN lattice.

Figure 3(c) shows representative CL kinetics profiles measured from a hBN flake that was coated with a grounded gold film (thickness ~7 nm). The electron beam energy and current were 5 keV and 1.6 nA, and $t_{sat}$ ~ 63 s and 50 s for the UV and B-centre emissions, respectively. For reference, typical corresponding values from uncoated hBN are much greater: $t_{sat}$ ~ 333 s and 142 s, respectively (see Figure 2(b,c)). The dramatic reduction in $t_{sat}$ (i.e., an increase in the CL kinetics rates) is consistent with an increase in $|\xi|$ and a corresponding increase in the electromigration rates of defects in hBN. The results in Figure 3(c) and (b) are therefore consistent, and both provide evidence for the role of $\xi$ and electromigration in the observed CL kinetics.

For completeness, we note that, whilst the directional trends in $t_{sat}$ shown in Figure 2 and 3 are representative of numerous measurements made during this study, the exact quantitative values can vary across any one flake of hBN by up to ~30% (under fixed beam parameters). This not surprising due to local variations in defect structure and density, which affect charging,[32,34,35] as well as the magnitude of $\xi$ and the rates of defect electromigration. In addition, the values of $t_{sat}$ were observed to vary from flake to flake due to both variations in flake thickness and flake quality (i.e., defect structure).

Summarizing the results in Figures 1-3, electron beam irradiation causes quenching of the UV emission at 305 nm, and the quenching is accompanied by generation and enhancement of the B-centre emission at 436 nm. The corresponding CL kinetics indicate that both of these processes are rate-limited by electromigration of defects in hBN. Specifically, charged defects migrate within the hBN lattice under the influence of an electric field produced due to charging caused by electron beam irradiation of hBN. The electromigration rate scales with $R$ and $|\xi|$, which in turn scales with the electron beam current and reciprocal electron beam energy. To optimize B-centre generation, the beam parameters should be selected to maximize both $R$ and $\xi$, accounting for the electron range $Z$ and thickness of the irradiated hBN. Most pertinently, thin flakes of hBN (< 10 nm) require the use of relatively low electron beam energies to ensure that both $R$ and $\xi$ are maximized – in general, the beam energy should be ~ 3 keV (lower beam energies are problematic due to contaminants at the sample surface, as is discuss below).

The CL kinetics also provide insights towards future identification of the atomic structure of the B-centre defect. Prior work has shown that B-centres can be generated by electron irradiation only in hBN that contains the UV emission with a ZPL at 305 nm (Figure 1(c)).[10] It was therefore tempting to suggest that the UV defects are converted to B-centres by the electron beam through a one-step pathways such as impact ionization of the UV defect, or cluster decomposition. However, the CL kinetics presented here indicate that this is not the case. Instead, the formation of B-centres involves the drift of a charged defect through the lattice, followed by the formation of the B-centre (e.g., via a process such as cluster formation or incorporation of an interstitial at a vacancy site).

Lastly, we observed that the B-centre CL kinetics can be affected profoundly by contaminants/residues that are present on the surface of hBN. The contaminants compromise the robustness and reliability of the electron beam B-centre generation method. We therefore demonstrate this artifact and present mitigation methods for B-centre engineering protocols.

## Contamination and background fluorescence

Hydrocarbon residues from hBN exfoliation and transfer methods are typically present on the surface of hBN and must be removed using hBN conditioning/cleaning procedures such as those employed in the present work (see Methods). However, the extent of contamination can vary substantially, and residual contaminants affect electron beam irradiation processes. An extreme example is shown in Figure 4(a), where the dark spot labeled "Vac" was produced using a 300 s irradiation. The spot is dark in the secondary electron image because the irradiation causes decomposition and immobilization of diffusing hydrocarbons, leading to the deposition of a carbon-rich film[19] on the surface of hBN. The spot diameter corresponds to the escape area of backscattered electrons from the surface[38] which is determined by the employed electron beam energy of 10 keV (and is much larger than the SEM electron beam diameter, which is on the order of nanometers). The dark streak above the spot indicates that the beam drifted over ~ 10 microns during the 300 s irradiation. CL kinetics measured during the irradiation are shown in Figure 4b. The curves do not show the typical, exponential decay and growth of the UV and B-centre emissions seen in Figures 2 and 3, and demonstrate almost complete failure of the electron beam B-centre fabrication technique.

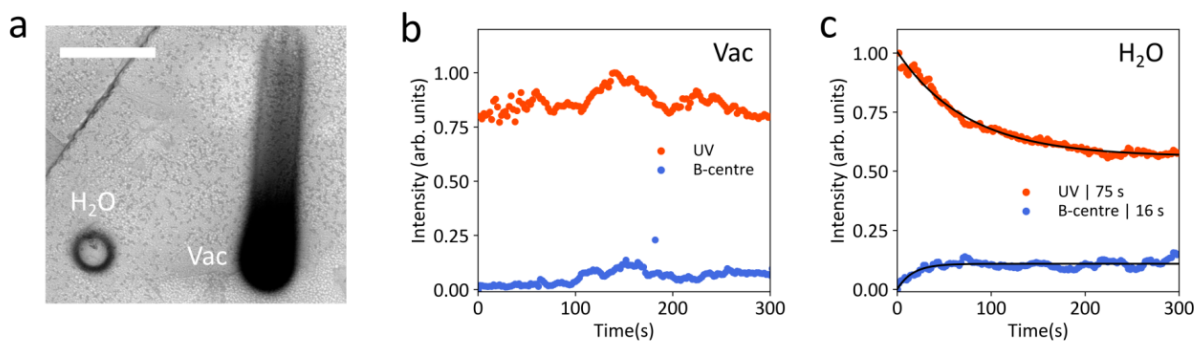

*Figure 4.* Role of hydrocarbon contaminants in CL kinetics. (a) SEM image showing two adjacent regions of hBN irradiated in high vacuum ("Vac", pressure ~ 3.2 x $10^{-6}$ mBar) and in the presence of $H_2O$ vapor (~1.5 x $10^{-5}$ mBar), using a 10 keV electron beam (1.6 nA, 300 s per irradiation). The scale bar is 5 µm. (b) UV and B-centre CL intensity measured versus time during the high vacuum electron beam irradiation. (c) UV and B-centre CL intensity measured versus time during the irradiation performed in the presence of $H_2O$. The sample contained an excessive amount of hydrocarbon contaminants which inhibited the B-centre electron beam generation process. The problem was mitigated by injecting $H_2O$ into the SEM chamber.

The deposition of carbon-rich films by an electron beam immobilization (cross-linking) of hydrocarbons is well-known in SEM.[19] The deposits modify charging behavior (and hence $\xi$), and can give rise to electron beam drift, as is illustrated by the extreme example in Figure 4(a). In addition, electrons lose energy to the deposits, thus reducing the energy per electron (and hence $\varepsilon$) dissipated in the underlying hBN. Deposit formation must therefore be minimized. This can be achieved using methods for the removal of hydrocarbons from the hBN surface, such as: (1) cleaning of the sample before it is transferred to the SEM (see Methods), (2) cleaning of the sample inside the SEM chamber using a low energy plasma that contains oxygen and volatilizes carbon,[39,40] and (3) performing electron beam irradiation in the presence of an oxygen-containing precursor gas.[38] The latter is demonstrated in figure 4(a). The dark halo labeled "$H_2O$" was produced using a 300 s irradiation in the presence of $H_2O$ vapor. The $H_2O$ was injected into the SEM chamber using the capillary-style gas injector shown in Figure 1(a), which increased the base pressure of the SEM chamber from 3.2 x $10^{-6}$ to 1.5 x $10^{-5}$ mBar. Surface-adsorbed $H_2O$ molecules give rise to dry chemical etching of carbon under the electron beam. This etching occurs concurrently and competes with carbon film deposition. Which process dominates (etching or deposition) is influenced by the electron flux, which decreases with increasing distance from the electron beam axis, yielding a clean hBN surface near the electron beam axis, surrounded by a carbon-rich halo near the periphery of the backscattered electron area[38] – i.e., the dark halo seen in Figure 4(a).

The $H_2O$ injection can have three dramatic effects on electron beam irradiation. First, it can suppress electron beam drift, as is demonstrated by Figure 4(a). Second, it can restore the exponential nature of the CL kinetics, as is demonstrated by Figure 4 (b) and (c). Third, it can increase the rates of CL kinetics – e.g., in Figure 4(c), $t_{sat}$ ~ 75 and 16 s for the UV and B-centre emissions, which are much smaller than expected from the trend in Figure 2(e) and (f), which shows that values greater than 410 and 128 s are expected at the employed beam energy of 10 keV. These three effects are expected consequences of a reduction in carbon film deposition, and an increase in $|\xi|$. The latter is expected because ionized $H_2O$ molecules are known to stabilize the surface potential of a dielectric under electron beam irradiation, in a manner similar to that of a grounded metallic coating.[34]

We note, however, that the example shown in Figure 4 is extreme in that the hBN sample was highly contaminated. It is used here to demonstrate clearly the effects of contaminants on electron beam irradiation of hBN. In addition, the effectiveness of the $H_2O$ injection technique is highly variable. Efficacy depends on the extent to which the hBN is contaminated, the $H_2O$ flow rate, and the electron beam irradiation parameters (including the beam energy and current). Moreover, if the carbon deposition rate is sufficiently low, electron beam irradiation in the presence of $H_2O$ gives rise to chemical dry etching of hBN,[41] which is problematic when irradiating thin flakes of hBN. Hence, a better solution is to clean hBN as best as possible prior to electron beam irradiation, perform the irradiation in high vacuum, tolerate a small amount of carbon buildup, and etch the carbon film *ex-situ* using UV ozone or an oxygen-containing plasma. We demonstrate this below, in the technologically-relevant limit of a short electron irradiation time (i.e., low electron dose) needed to generate single B-centre defects (or small ensembles).

We explored a number of protocols to suppress hydrocarbon contaminants (see Figure S1), all based on chemical dry etching of carbon. The most effective of these is shown schematically in Figure 5(a). Exfoliated hBN flakes were first annealed in air for 2 hours at 500 °C prior to transfer into an SEM chamber. The flakes and the SEM chamber were then irradiated for 20 min by a low pressure radiofrequency plasma, using air as the plasma precursor gas.[39,40] Multiple electron beam irradiations were then performed in high vacuum to generate arrays of B-centres as a function of electron irradiation time (i.e., electron dose). Each irradiation generated one or more B-centres, and caused the deposition of a small amount of carbon on the surface of hBN. The carbon was removed *ex-situ* using a 30 min UV ozone treatment. The arrays were characterized by confocal PL mapping and spectroscopy. CL was not appropriate in this instance for a number of reasons. First, we used low electron irradiation doses, and the CL system collection efficiency is much lower than that of the confocal PL setup. Second, we had to prevent electron beam modification of hBN during fluorescence mapping and spectroscopy. Third, fluorescence from the carbon film was not visible in CL spectra, as is discussed below.

Confocal PL maps of a 5 x 5 irradiation array collected using an excitation laser of 405 nm before and after ozone processing are shown in Figure 5(b) and (c), respectively. The diameter of each spot in the array was reduced by the ozone treatment. The initial large spots correspond to carbon deposited by the electron beam (analogous to the spot labeled "Vac" in Figure 4(a)), which is fluorescent and overlaps spectrally with the B-centre emission. This is demonstrated by the PL spectra in Figure 5(d), collected from one spot in the array before and after ozone processing. The spectra reveal a broad background emission centered on ~ 550 nm that overlaps with the B-centre emission. The background is problematic as it reduces the B-centre signal-to-background ratio, and hence the purity of photons emitted by B-centre quantum emitters. The ozone treatment suppresses the background fluorescence substantially due to chemical dry etching of carbon present on the surface of hBN. We note that the background emission is not visible in CL spectra (Figures 1 and 2), likely due to competitive recombination, [42] and a low fluorescence efficiency of the carbon film relative to the intense UV and B-center ensemble emissions studied in CL experiments. We also note that the films can not explain the exponential decay and growth of the UV and B-centre CL emissions.

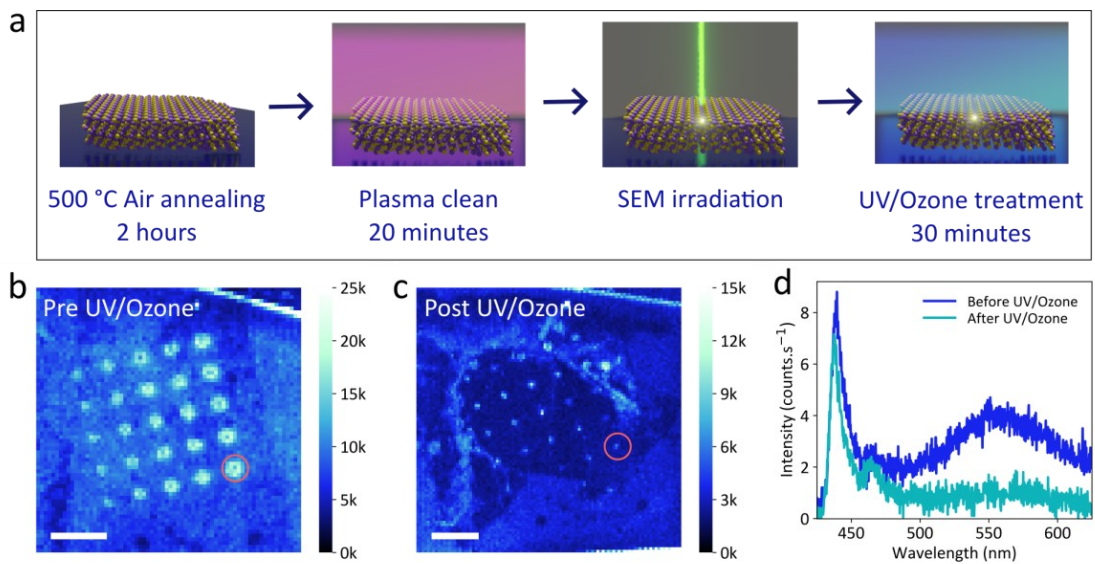

**Figure 5.** *Suppression of background fluorescence by UV ozone processing. (a) Protocol used to generate B-centres. Exfoliated hBN was annealed in air prior to transfer to an SEM chamber where it was plasma cleaned before irradiation in high vacuum by an electron beam. Subsequently, an ex-situ UV ozone treatment was used to dry etch carbon and suppress background fluorescence. (b) Confocal PL map collected before ozone exposure, showing a 5 x 5 array of B-centres generated by an electron beam (energy = 3 keV, current = 1.6 nA, five doses in the range of $1.0 \times 10^{10}$ to $2.5 \times 10^{11}$ electrons per spot, repeated five times). (c) PL map of the same array after a 30 minute UV ozone treatment. The treatment caused a reduction in the diameter of each spot in the array. (d) PL spectra from the spot circled in (a) and (b) acquired before (teal) and after (blue) UV ozone exposure, showing suppression of a broad background emission centered on ~550 nm. The scale bars in (b) and (c) designate 10 μm.*

The hBN pre- and post-processing protocol shown in Figure 5(a) greatly improves the robustness and efficacy of the electron beam B-centre generation method. It can be used to generate single B-centre quantum emitters with high photon purity in relatively thick (~ 100 nm) hBN, and also quantum emitters in very thin (<10 nm) flakes of hBN, as is demonstrated in the Supporting Information, Figures S2 and S3, respectively. We note however, that B-centres in very thin flakes are optically unstable, as is demonstrated in Figure S3 of the Supporting Information. Such instabilities have been noted previously for quantum emitters in monolayers and thin flakes of hBN.[43,44] Mitigation of the instabilities is a logical next step, but it is beyond the scope of the present work.

# Conclusion

Time-resolved CL spectroscopy was used to study the generation of B-centre quantum emitters by electron beam irradiation of hBN. The emitter generation rate is limited by electromigration of charged defects in the hBN lattice. The electromigration is caused by an electric field produced due to charging of hBN (a wide bandgap dielectric). The defect migration rate scales with the field intensity and the energy per unit volume per unit time dissipated in hBN by the electron beam. To optimize B-centre generation, the beam energy must be selected to maximize the electromigration rate, whilst minimizing the deposition of carbonaceous films on the surface of hBN. The latter is an artifact caused by electron beam irradiation of hydrocarbon contaminants produced during exfoliation, transfer and handling of hBN. The carbonaceous films interfere with B-centre generation and produce a broadband background PL emission that limits the photon purity of B-centre quantum emitters. These problems are particularly acute for thin flakes of hBN (< 10 nm), where they can not only compromise, but even prevent B-centre generation entirely. To solve these issues, we employed a beam energy of ~ 3 keV (to optimize electromigration) in conjunction with hBN pre- and post-processing treatments that volatilize carbon-containing molecules and films at the hBN surface. Using this approach, we achieved B-centre quantum emitters in hBN flakes as thin as 8 nm.

# Methods

### Preparation of samples for in-situ CL experiments

Silicon substrates were cleaned using sequential steps of ultrasonication in acetone and isopropyl alcohol. The substrates were then dried under flowing nitrogen, and exposed to UV ozone for 30 minutes.[45] Carbon-doped hexagonal boron nitride (hBN) bulk crystals, sourced from the National Institute for Materials Science (NIMS), were exfoliated onto the cleaned substrates using the scotch tape method. Flake thicknesses were assessed using an optical microscope, according to flake colour.[46] Some samples (Figure 3c) were coated with 7 nm of gold using a Leica EM ACE600 high vacuum sputter coater. The gold coating and the Si substrate were electrically grounded. All samples were annealed in air at 500 °C for 12 hours prior to transfer to an SEM chamber where they were cleaned further for 30 min using an air plasma cleaner.[39,40]

### Preparation of samples characterized by PL spectroscopy

Si substrate cleaning hBN exfoliation was done as is detailed above. The samples were then annealed in air for 2 hours at 500 °C prior to transfer to an SEM chamber where they were cleaned further for 20 min using an air plasma cleaner.[39,40] Electron beam irradiation was performed using an electron beam energy of 3 keV and a current of 1.6 nA. Following irradiation, samples were exposed ex-situ to ozone for 30 min to remove carbon films deposited on the surface of hBN by the electron beam. The thickness of hBN flakes was confirmed using a Park XE7 atomic force microscope.

### Electron beam irradiation and cathodoluminescence analysis

Electron beam irradiation was performed using a ThermoFisher G4 dualbeam microscope. Cathodoluminescence was collected using a ball lens coupled to an optical fibre connected to a QE Pro and a QE65000 Ocean Optics spectrometer located outside the SEM vacuum chamber.

### Photoluminescence

Photoluminescence measurements were carried out using a home-built confocal microscope with a 405 nm continuous wave laser (PiL040X, A.L.S. GmbH). The laser power was measured before a Nikon TU Plan Fluor 100x/0.90 NA objective lens. PL mapping and spectral acquisition were performed at 200 µW using 460 ± 30 band-pass filters and 430 long-pass filters, respectively. Photons were collected using avalanche photodiode single-photon detectors (APDs) (Excelitas Technologies) or a spectrometer (Princeton Instruments, Inc). Second-order correlation measurements were performed using a 50/50 beam-splitting fibre and a Picoharp 300 correlation unit from Picoquant.

## Acknowledgements

This research is supported by the Australian Research Council (CE200100010, FT220100053) and the Office of Naval Research Global (N62909-22-1-2028). The authors thank the ANFF node of UTS for access to facilities. We acknowledge Takashi Taniguchi (the National Institute for Materials Science (NIMS)) for providing hBN crystal.